\documentclass
[aps,superscriptaddress,prd,
onecolumn,
floatfix,
amsmath,amssymb,amsfonts]{revtex4-2}%

\usepackage[pdftex]{graphicx}
\usepackage{color}
\usepackage[colorlinks,linkcolor=blue,anchorcolor=violet,citecolor=red]{hyperref}
\usepackage{float}
\usepackage{mathrsfs}

\begin{document}
\title{Expanding Edges of
 Quantum Hall Systems
 in a Cosmology Language \\ 
 -- Hawking Radiation from de Sitter Horizon in Edge Modes}
\author{Masahiro Hotta}
\email{hotta@tuhep.phys.tohoku.ac.jp}
\affiliation{Department of Physics, Tohoku  University, Sendai 980-8578, Japan}

\author{Yasusada Nambu}
\affiliation{Graduate School of Science, Nagoya University, Nagoya 464-8601, Japan}
\email{nambu@gravity.phys.nagoya-u.ac.jp}

\author{Yuuki~Sugiyama}
\affiliation{Department of Physics, Kyushu University, 744 Motooka, Nishi-Ku, Fukuoka 819-0395, Japan}
\email{sugiyama.yuki@phys.kyushu-u.ac.jp}

\author{Kazuhiro Yamamoto}
\affiliation{Department of Physics, Kyushu University, 744 Motooka, Nishi-Ku, Fukuoka 819-0395, Japan}
\affiliation{Research Center for Advanced Particle Physics, Kyushu University, 744 Motooka, Nishi-ku, Fukuoka 819-0395, Japan}
\email{yamamoto@phys.kyushu-u.ac.jp}

\author{Go Yusa}
\email{yusa@tohoku.ac.jp}
\affiliation{Department of Physics, Tohoku University, Sendai 980-8578, Japan}
\affiliation{ 
Center for Spintronics Research Network, Tohoku University, Sendai 980-8577, Japan}
\date{\today}

\begin{abstract}
Expanding edge experiments are promising to open new physics windows of quantum Hall systems.  
In a static edge, the edge excitation, which is described by free fields decoupled with the bulk dynamics, is gapless, and the dynamics preserve conformal symmetry. 
When the edge expands, such properties need not be preserved. We formulate a quantum field theory in 1+1 dimensional curved spacetimes to analyze the edge dynamics. We propose methods to address the following questions using edge waveforms from the expanding region: 
Does the conformal symmetry survive? Is the nonlinear interaction of the edge excitations induced by edge expansion?
 Do the edge excitations interact with the bulk excitations? 
We additionally show that the expanding edges can be regarded as expanding universe simulators of two-dimensional dilaton-gravity models, including the Jackiw-Teitelboim gravity model. As an application, we point out that our theoretical setup might simulate emission of analog Hawking radiation with the Gibbons-Hawking temperature from the future de Sitter horizon formed in the expanding edge region.
\end{abstract}
\maketitle
\section{Introduction}

How did our universe develop from its earliest moments? To answer this major
question of quantum cosmology, great efforts have been made
both in theory and in the attempts of verifying the theory by observations of the cosmic
background radiation and black holes. An ideal scenario would be the reproduction of the origin and
evolution of the universe or producing black holes in a laboratory and thereby
experimentally verifying the theory through controlling key parameters. Regarding this,
analog experiments have been performed in various systems \cite{garay,schutzhold,rousseaux,weinfurtner}.

Quantum Hall (QH) systems are unique and promising. A
QH system \cite{vk, tsui}\ emerges when a strong perpendicular magnetic
field ($B$) is applied to two-dimensional (2D) electrons when the Landau
level filling factor, $\nu =2\pi \hbar n_{e}/(eB)$, 
becomes an integer or a rational fraction, where $\hbar$ and $e$ are the reduced Planck constant and the elementary charge, respectively.

The theoretical study of QH systems has contributed to modern
understanding of condensed matter physics. Notably, the QH
systems are regarded as typical topological materials consisting of the bulk
and edge. The dynamics in the bulk yield a large energy gap in its
dispersion relations. In the bulk, there exist various gapped excitations
such as magneto-rotons \cite{girvin,pinczuk,kukushkin,kamiyama} and anyonic quasiparticles \cite{leinaas,nakamura,bartolomei}. The dispersion relation
of the edge currents are firmly protected owing to the topological structure of
the systems, and the edge excitations are always in gapless modes. The
effective theory of the bulk is given by a topological field theory
referred to as the Chern-Simons gauge theory \cite{CS1,CS2,CS3,CS4,CS5}. In the topological field theory,
no local dynamics appears, and properties of the system are stable under
local perturbations in the bulk. The bulk action is not invariant under the gauge transformation at the edge. This unsatisfactory gauge-variant behavior of the bulk is compensated by adding a gauge-variant edge action with
quantum anomaly. Hence, this predicts the edge dynamics \cite{wen}. It is
possible to express the same edge degrees of freedom by both a fermionic field
and bosonic field via the statistics transmutation in one-dimensional space.
It is known that the bosonic field corresponds to the charge density of the edge
current, and can be directly observed by measuring the voltage deviation in the
edge experiments \cite{kamiyama,m1,m2,m3,m4,m5,m6}. The edge
effective theories are given by free field theories with a chiral condition.
Here, the chiral condition implies that the excitations move in one direction along
the edge and do not return to the upstream region. The theories belong to
a class of conformal field theory (CFT) in 1+1 spacetime dimensions, which
preserves an infinite dimensional conformal symmetry, referred to as the Virasoro
symmetry \cite{V}. The total gauge symmetry of the bulk-edge composite system ensures
the gapless property of the edge excitations.

Until now, all experiments of QH systems have been performed in a static situation, except for local edge excitations \cite{kamiyama,m1,m2,m3,m4,m5}. The electrons are confined in the bulk region by the static
electric field created by surface potential of host semiconductors of the 2D
electrons. Thus, the edge attached to the bulk remains unchanged in time. In reference \cite{HMY}, expanding edges were proposed. The edge expands by
gradually relaxing the external electric fields through continuous electron
supply into the bulk. The excitations moving along the edge are affected by
the expansion.  In such situations, it is quite nontrivial that the conformal symmetry still holds. Recall that the static conformal symmetry is protected by topological properties of QH systems which are insensitive to
microscopic details. It is known that such topologically robust phenomena are not limited to static systems but
may also emerge in some periodically-driven quantum systems  \cite{Kitagawa}. From this viewpoint, it can be expected that the conformal symmetry still survives in the expanding QH edges. On the other hand, there exists a risk of terminating the conformal symmetry in the realistic experiments. It is considered that the conformal symmetry does not survive in some 1+1 dimensional acoustic systems described by minimally-coupled massless scalar field \cite{BLV}. Also the conformal symmetry can be broken by inhomogeneity induced by impurities and/or time-dependence of the realistic QH systems. 
About the effective field theory for expanding edge systems, it should be stressed that answers to the following fundamental questions are unknown at present:
\begin{enumerate}
\item Does the edge current remain gapless in the expansion? Specifically,
is the conformal symmetry preserved at the boundary?

\item Beyond the free field theory, does nontrivial interaction among edge
excitations emerge in the expansion?

\item During the expansion, do excitations in the edge interact with gapped
excitations in the bulk?
\end{enumerate}

The expanding edge experiments will provide crucial results for the answers to the above questions. Note that the experiments are able to achieve this aim even in classical regimes of the systems. Thus the experiments are quite promising to reveal new physics of the systems under such dynamical backgrounds.

 In quantum regimes with extremely low noise, QH edges may be applied to future development  in quantum information science. The edge excitations transport quantum states, i.e., quantum information along an edge connected  between distant points inside a quantum device. This means that the QH edges play a role of quantum channels. In such situations, the expanding region provides a useful quantum gate referred to as quantum squeezing for the quantum states \cite{HMY}. In addition, measurement of zero-point fluctuation at the downstream of the edge and its feedback to the upstream are capable of extracting local zero-point energy at the upstream accompanied by generation of a negative energy density region in the edge \cite{m6}. The protocols are referred to as  quantum energy teleportation (QET) \cite{QET1,QET2,QET3}. The inevitable energy cost $E_\text{in}$ of the measurement at the downstream is regarded as the input energy  of QET, and the extracted positive energy $E_\text{out}$ at the upstream is regarded as the output energy of QET satisfying $0<E_\text{out}<E_\text{in}$.  It is expected that long-distance QET may be attained utilizing quantum entanglement of the expanding edge systems \cite{HMY}.

The proposed experiment of expanding edges in \cite{HMY} will also have
large implications to cosmology. The expanding edge can be regarded as a
simulator for the expanding universes in 1+1 spacetime dimensions. In particular, the
charge density of the edge current is capable of playing the analog role of a
quantum field in curved spacetimes. In curved spacetime field theories, many
intriguing phenomena are predicted \cite{BD}. One of these
phenomena is the Hawking radiation emission out of black hole horizons \cite%
{hawking}. The black hole evaporation induced by Hawking radiation has
not yet been confirmed in astrophysical observations, but many believe that
black holes are evaporating. If a black hole completely evaporated and its
Hawking radiation remained in space, the thermal radiation would be in a
mixed state with a single parameter, that is, its temperature. Thus, at a
glance, the radiation does not appear to carry the huge amount of information originally
stored inside black holes. Where has the information gone? \ Suppose that the
initial state of the black hole is a pure state and that the evolution to the mixed
state is realized resulting in a loss of quantum coherence without any environmental interaction. Then unitarity would be broken, which is one of fundamental
laws of quantum mechanics. This unresolved problem is referred to as the
information loss problem.

In the exploration of information loss, crucial missing links of Hawking’s original analysis are known \cite{H2}. For example, his
calculation is only semi-classical. The matter is quantized, but the
spacetime remains classical in his analysis. Thus, his theory does not cover
the final burst of black holes. In the theory of general relativity for classical
spacetime, a curvature singularity appears and loses predictability by the
theory after the burst. To avoid this flaw, quantum theory is
required, which can treat quantum spacetimes with no singularities. Although several candidates such as string theory exist,
quantum gravity theory remains elusive owing to the lack of strong guiding
principles, actual measurements, and convincing observations to determine
the theory uniquely. To resolve the information loss, any
findings and inspiration, which are delivered by analog black
hole experiments in condensed matter physics  including QH systems, are considered useful \cite{S}.

Such analog experiments also shed light on the trans-Planckian problem 
\cite{tP}. The typical wavelength of a thermal quantum particle emitted from a black
hole with mass $M$ is computed as $\lambda =O\left( GM/c^2\right) $, using the 
Hawking temperature $T=\hbar c^3 /(8\pi k_{B}GM)$ \cite{hawking}. Here, $c$ is the velocity of light, $G$ is the
gravitational constant, and $k_{B}\,$\ is the Boltzmann constant. The particle
experiences severe red shift by gravitational potential when it propagates
to spatial infinity. At the previous point near the horizon where the particle is
created, the wavelength of the particle mode is much shorter than the Planck
length $l_{P}=\sqrt{G\hbar/c^3}$. Thus, a precise description of the mode in
the past regime is required for fundamental microscopic theories such as
string theory. It is possible that Hawking’s approximation is
incorrect and has serious discrepancy. However, his
results appear to be correct. In reality, the formula for Hawking
temperature $T$ and derived entropy $S=k_{B} c^3 \mathscr{A}/(4G\hbar )~$for
black holes with horizon area $\mathscr{A}$ are consistent with the other
theoretical results of generalized thermodynamics \cite{GTD1, GTD2}
and statistical mechanics with state counting in string theory \cite{SV}.
Thus, why Hawking's analysis works so well despite being semi-classical is a mystery. This is called the trans-Planckian problem.

We can reconsider the trans-Planckian problem in condensed matter
physics.  If Hawking's prediction is correct, analog black holes also emit Hawking radiation, no matter what the analog black holes are made of \cite{U}. Analog systems corresponding to black holes possess natural cutoff
length, just like the magnetic length $l_{B}=\sqrt{\hbar/(eB)}$ of\
electron in QH systems. This phenomenon poses an analog version of the trans-Planckian problem by regarding $l_{B}$ as $l_{P}$. Thus, when the original problem is considered, it is 
significant to analyze the analog problem. Notably, the trans-Planckian problem also appears in an inflationary
expanding universe owing to the existence of the cosmological horizon in the
universe. Hawking radiation is emitted out of the horizon and provides
nondecaying thermal fluctuation even in the extreme expansion of the universe. Therefore,
expanding edge experiments using QH systems are also appropriate for
exploration of the corresponding trans-Planckian problems associated with expanding universe.

In the inflationary universe, there exists another conceptual issue related to quantum field theory in the expanding universe. The accelerated expansion in the inflationary universe provides large-scale quantum fluctuations of the inflaton field over the Hubble horizon scale. This primordial fluctuation leads to gravitational instabilities that ultimately form large-scale structures in our universe  \cite{KT}. 
However, it is not known how the  quantum-classical transition of field fluctuation occurs in
a long wavelength region beyond the Hubble horizon scale. Because the details and mechanisms of the transition process are not known, it remains far from profound understanding. In the analysis, quantum entanglement is known to be
capable of capturing the quantum-classical transition \cite{N1,N2,N3,N4}.  The bipartite entanglement between two spatial regions will be lost in the course of evolution \cite{N1,N4}, and entanglement harvesting from spatially separated regions becomes impossible \cite{N2,N3}. Additionally a study of entanglement structure in quantum states in de Sitter space \cite{HY} may be useful to resolve the issue. It is possible to simulate the generation of analog primordial fluctuations experimentally, using expanding analog universes of QH systems. We are able to detect and study the entanglement of long-wavelength quantum fluctuations in the expanding universe. This direction of analog experiments in QH systems facilitates us to capture a detailed understanding of the quantum-classical transition process in the early universe. 
\begin{figure}[h]
\centering
\includegraphics[width=0.6\linewidth]{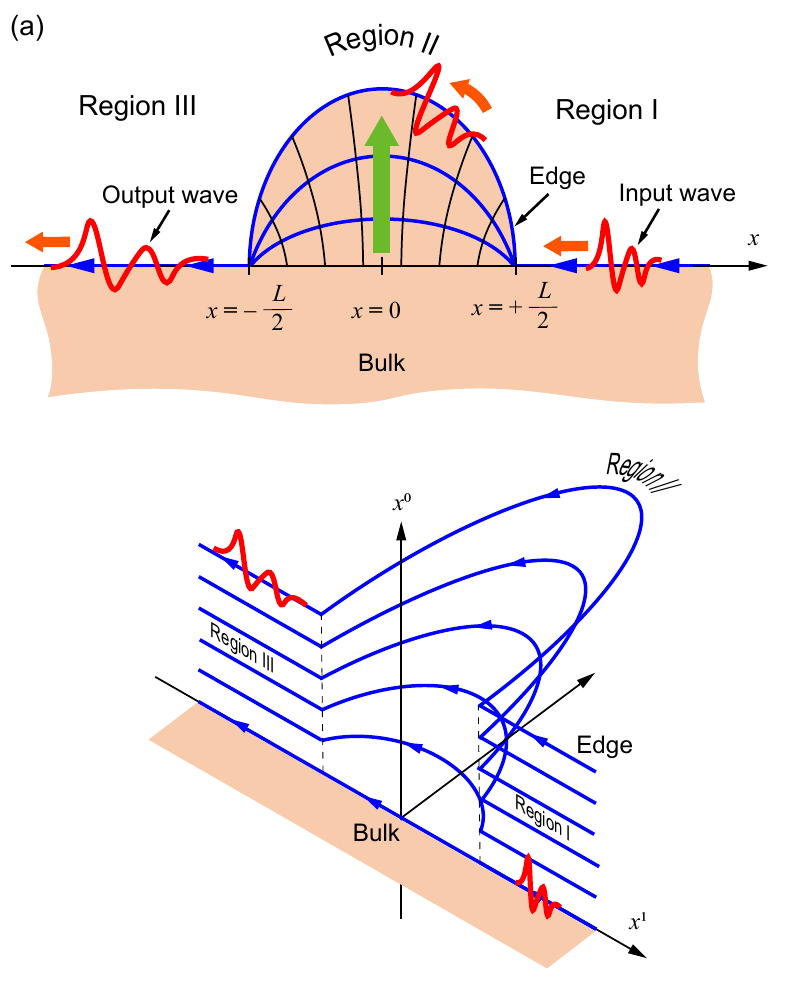}
\caption{(Color online) Schematic of setup. (a) Top view of the QH system consisting of the bulk and the edge. Region I and Region III are flat and Region II is the expanding region. (b) The QH edge in 2D spacetime using the coordinates $x^0=vt$ and $x^1=x$.}%
\label{fig:fig1}%
\end{figure}

In this paper, we formulate a general theory of the expanding edge of QH systems in a cosmology language, i.e., quantum field theory in a curved spacetime. Taking account of the great success of effective field theory of static edge systems, we adopt the similar description of effective field theory for the time-dependent edge systems. It is assumed that such an effective theory description is valid as long as we consider a long wavelength regime. In fact, the QH edge experiments usually observe edge excitation wavelengths about 100 times larger than the magnetic length $l_{B}$ \cite{kamiyama} \cite{m5}. Hence the effective theory description may be suitable for the realistic experiments.  The derivation of the effective equations obeyed by edge modes from a microscopic Hamiltonian of the dynamical QH system is out of our aim in this paper, and will be reported in future.  
The theory will be directly applied to analyses of future experiments of the edge expansion, and stimulate an interdisciplinary interest connecting condensed matter physics and cosmology.
First, we divide the entire edge into three regions, as depicted in Fig.~\ref{fig:fig1}: input flat Region I,
expanding Region II, and output flat Region III. Region I is defined as
$x>+L/2$. Region II is defined as $-L/2\leq x\leq +L/2$. Region 
III is defined as $x<-L/2$. Here, $L$ denotes the width of  Region II when
the edge is static. Edge excitation starts from Region I and runs
through the expanding Region II, and finally enters  Region III.

The structure of this paper is as follows.
In Sec. II, conformally flat coordinates $x^{\pm }$ of general relativity
are introduced for the expanding Region II. Additionally, we define another
conformally flat set of coordinates $x_\text{out}^{\pm }$ of the expanding Region II,
which are smoothly connected with the conformally flat coordinates of the output
Region III. Similarly, we define different conformally flat coordinates $%
x_\text{in}^{\pm }$ of the expanding Region II, which are smoothly connected
with the conformally flat coordinates of the input Region II. The coordinate
transformation is derived between $x_\text{out}^{\pm }$ and $x_\text{in}^{\pm }$. We
introduce Bogoliubov coefficients between in-mode functions for Region I and
out-mode functions for Region III. The implication of the Bogoliubov coefficients
measurement in realistic experiments is discussed for future experiments.
 In Sec. III, we show that the expanding edges can be regarded as simulators of homogeneous and isotropic universes referred to as Friedmann--Lema\^{\i}tre--Robertson--Walker (FLRW)\ universes in 1+1 dimensions.  The dynamics are described by 2D gravity models with a scalar field $\Phi$, which is called dilaton field. The models include the famous Jackiw--Teitelboim (JT) model with a
negative constant curvature $R=-4\lambda ^{2}$ \cite{T, J}. The JT model
has recently attracted considerable attention because the gravity model is closely
connected with a many-body model called the Sachdev-Ye-Kitaev (SYK) model \cite%
{SY,K,SYK}, and it suggests a nontrivial example of an anti-de Sitter space (AdS)/CFT correspondence \cite{M}. To connect the edge system with the AdS/CFT correspondence, the three-dimensional AdS chiral gravity may be interesting, which is dual to 2D CFT  \cite{LSS}.  Kerr/CFT correspondence may be also related to the edge dynamics \cite{KAdS1,KAdS2,KAdS3}.
The JT model with positive cosmological constant is referred to as the de Sitter JT (dSJT) model in this paper. The dSJT model achieves an exponentially fast expansion of the analog universe. Thus, the rapidly expanding edge experiments are quite fascinating for the study of the trans-Planckian problem and the quantum-classical transition problem in the dSJT model. 
In Sec. IV, as an application of our experimental setup, we present  the analogue Hawking radiation from Region II. With the de Sitter type expanding edge region, a future  horizon is formed and the power of signals measured in Region III shows the Planckian distribution with Hawking temperature $\lambda/(\sqrt{2}\pi)$.   
We summarize our results in Sec. V. In the Appendix, useful formulas are
attached for 2D gravity.

\section{Conformally Flat Coordinates for Expanding region}

Let us first consider static edge systems. First, we define the edge excitation
as the left mover with respect to the spatial coordinate $x$ at the edge. Specifically,
excitation runs from the right region with $x>0$ to the left region with 
$x<0$. Subsequently, $v$ denotes the edge current velocity of the system. Introducing relativity notations, as in cosmology, is useful. The time $t$ is
replaced by $x^{0}=vt$. The spatial coordinate $x$ is denoted by $x^{1}$, and light cone coordinates are then defined as $x^{\pm }=x^{0}\pm x^{1}$, even though 
$v$ is not the actual velocity of the light. The edge excitation is represented by a
free real field $\varphi (x^{+})$. The field $\varphi (x^{+})$ is
proportional to the voltage fluctuation on the edge and can be directly measured
in experiments \cite{Yoshioka}. In this situation, the spacetime for the edge is just a flat
spacetime called the Minkowski spacetime. The square of the spacetime-invariant
distance, which is called metric form, is given by 
\begin{equation}
ds^{2}=-v^{2}dt^{2}+dx^{2}=-\left( dx^{0}\right) ^{2}+\left( dx^{1}\right)
^{2}=-dx^{+}dx^{-}.
\end{equation}%
The aforementioned equation fixes the metric matrix for each coordinate system as 
\begin{equation*}
\begin{bmatrix}
g_{00} & g_{01} \\ 
g_{10} & g_{11}%
\end{bmatrix}%
 =
\begin{bmatrix}
-1 & 0 \\ 
0 & 1%
\end{bmatrix}%
,\quad
\begin{bmatrix}
g_{++} & g_{+-} \\ 
g_{-+} & g_{--}%
\end{bmatrix}%
=
\begin{bmatrix}
0 & -1/2 \\ 
-1/2 & 0%
\end{bmatrix}%
,
\end{equation*}%
where $ ds^{2}=g_{\mu \nu }dx^{\mu }dx^{\nu }$. 
Note that $\varphi (x^{+})$ satisfies the massless Klein-Gordon equation, 
\begin{equation}
\left[-\left(\partial_0\right) ^{2}+\left(
\partial_1\right) ^{2}\right]\varphi =-4\partial_+\partial_-\varphi =0,
\end{equation}
and the chirality condition $
\partial_{-}\varphi =0$
implying that the edge current moves only in the left direction.

Let Region II start expanding uniformly at $t=0$, with the remaining Regions I and III  being unchanged. The spacetime metric in Region II is described by the
following equation:%
\begin{equation}
ds^{2}=-v^{2}d\tau ^{2}+a^2(\tau )dx^{2},  \label{e1}
\end{equation}%
where $\tau $ is the proper time for an observer at $x=const.$, and it is equal to $%
t $ when $t<0$. The positive function $a(\tau )$ of $\tau $ satisfies $%
a(\tau )=1$ when $\tau <0$. At a fixed time $\tau $, the physical distance $l$
between two points at $x=x_{1}$ and $x=x_{2}~ (>x_{1})$ is computed as $%
l=a(\tau )(x_{2}-x_{1})$. Thus, $a(\tau )$ determines the real size of the
spatial region using the coordinate values, and is called the scale factor of
the expanding universe. Let us rewrite Eq.~(\ref{e1}) as 
\begin{equation}
ds^{2}=a^2(\tau )\left[ -v^{2}\left(\frac{d\tau}{a(\tau)}\right)
^{2}+dx^{2}\right].
\end{equation}%
By defining the relativistic coordinates as 
\begin{equation}
x^{0}=v\,t=v\int_{0}^{\tau }\frac{d\tau ^{\prime }}{a(\tau ^{\prime })}%
,\quad x^{1}=x,
\end{equation}%
and introducing a real function $\Theta $ of $x^{0}$, which obeys $\Theta
( x^{0}) =\ln a(\tau )$, we obtain the following metric form in
the light cone coordinates:%
\begin{equation}
ds^{2}=-\exp \left[2\Theta\!\left(\frac{x^{+}+x^{-}}{2}\right) \right]
dx^{+}dx^{-},  \label{e2}
\end{equation}%
where $x^{0}=(x^{+}+x^{-})/2$. The form of Eq.~(\ref{e2}) is defined as
conformally flat. 

Note that, at least in a local spacetime region,
any metric form in 1+1 dimensions can be rewritten using a general
coordinate transformation to the conformally flat form 
$
ds^{2}=-\exp \left( 2\Theta( x^{+},x^{-}) \right) dx^{+}dx^{-}
$
where $\Theta( x^{+},x^{-})$ is a real function of the light
cone coordinates $x^{\pm }$. This is because we have two degrees of freedom as a general coordinate transformation: $x'^+=f_+ (x^+ ,x^-)$ and $x'^- =f_- (x^+ ,x^-)$. By considering the two functions $f_\pm$ in the appropriate forms, the two conditions $g'_{++}=0$ and $g'_{--}=0$ are satisfied in the new coordinates ${x'}^{\pm}$. Hence, we always obtain conformally flat metric forms.  
The factor $\exp \left( 2\Theta \left(
x^{+},x^{-}\right) \right) $ is called conformal factor.

\subsection{Conformally Flat Coordinates Connecting to the Expanding region and Output region}

The conformally flat coordinates $x^{\pm }$ in Eq.~(\ref{e2}) as defined in
Region II cannot extend to Region III as the conformal factor jumps
at the boundary $x=x^1=-L/2$. However, another set of conformally
flat coordinates $x_\text{out}^{\pm }=x_\text{out}^{0}\pm x_\text{out}^{1}$ can be introduced, which smoothly
connect Region II and Region III at the boundary $x=x_{1}=-L/2$.
To ensure that $x_\text{out}^{\pm}$ are conformally flat coordinate systems, the original coordinates $x^{\pm}$ must be $x_\text{out}^{\mp}$-independent functions, i.e., $x^{\pm}=x^{\pm}\!\left( x_\text{out}^{\pm}\right) $.  Then, the metric form in Eq.~(\ref{e2}) is
given by the new coordinates $x_\text{out}^{\pm }$ as 
\begin{equation}
ds^{2}=-\exp \left[ 2\Theta\!\left(\frac{x^{+}\!\left( x_\text{out}^{+}\right)
+x^{-}\!\left( x^-_\text{out}\right) }{2}\right) -\ln\left( \frac{dx_\text{out}^{+}%
}{dx^{+}}\,\frac{dx^-_\text{out}}{dx^{-}}\right) \right]
dx_\text{out}^{+}\,dx_\text{out}^-.
\end{equation}%
The above metric form remains the flat metric form in Region III as 
$ds^{2}=-dx_\text{out}^{+}\,dx_\text{out}^{-}$. 
The coordinate transformation, $x^{\pm }=x^{\pm }\!\left( x_\text{out}^{\pm
}\right) $, is uniquely determined by the following two conditions:
\begin{itemize}
\item[(i)] The spatial position of the boundary in the new coordinates is
given by $x^{1}=x_\text{out}^{1}=-L/2$;
\item[(ii)] The time coordinates at the boundary coincide with each other: $%
x^{0}=x_\text{out}^{0}$.
\end{itemize}
Here, conditions (i) and (ii) can be replaced by (i) and a conformal factor continuity condition, i.e., the conformal factor in $x_\text{out}^{\pm}$ is continuous at the boundary between Region II and Region III such that
\begin{itemize}
\item[(ii)']  
\begin{equation}
1
=\exp\left( 2\Theta\!\left( \frac{x^{+}\!\left( x_\text{out}^{+}\right)
+x^{-}\!\left( x_\text{out}^{-}\right) }{2}\right) 
 -\ln \left( \frac{dx_\text{out}^{+}%
}{dx^{+}}\,\frac{dx_\text{out}^{-}}{dx^{-}}\right) \right)\quad\text{at~}x^1=-L/2.
\end{equation}
\end{itemize}
The condition (i) yields the following condition:%
\begin{equation}
x_\text{out}^{+}\!\left(x^{0}-\frac{L}{2}\right)
-x_\text{out}^{-}\!\left(x^{0}+\frac{L}{2}\right) =-L.
\label{e3}
\end{equation}%
By taking the derivative with respect to $x^{0}$, we get the following relation:
\begin{equation}
\frac{dx_\text{out}^{+}}{dx^{+}}\!\left(x^{0}-\frac{L}{2}
\right) =\frac{dx_\text{out}^{-}}{dx^{-}}\!\left(x^{0}+\frac{L%
}{2}\right) .  \label{e4}
\end{equation}%
Using the condition (ii)', the following relation is derived:
\begin{equation}
\frac{dx_\text{out}^{+}}{dx^{+}}\!\left(x^{0}-\frac{L}{2}
\right) \frac{dx_\text{out}^{-}}{dx^{-}}\!\left(x^{0}+\frac{L}{%
2} \right) 
=\exp \left(2\Theta(x^{0}) \right) .
\label{e5}
\end{equation}%
From the above equation and Eq.~(\ref{e4}), $x_\text{out}^{+}$ obeys the following
relation:
\begin{equation}
\frac{dx_\text{out}^{+}}{dx^{+}}\!\left(x^{0}-\frac{L}{2}
\right) =\exp \left( \Theta( x^{0}) \right) .
\end{equation}%
By changing the free parameter $x^{0}$ to $x^{+}=x^{0}-%
L/2 $, this equation can be replaced by
\begin{equation}
\frac{dx_\text{out}^{+}}{dx^{+}}\!\left( x^{+}\right) =\exp \left( \Theta \!\left(
x^{+}+\frac{L}{2}\right) \right) .
\end{equation}%
Integration of this equation yields
\begin{equation}
x_\text{out}^{+}(x^{+})=\int_{0}^{x^{+}}dy\exp \left( \Theta\!\left( y+\frac{L}{2}%
\right) \right),  \label{e8}
\end{equation}
where the integration constant is fixed, such that $x_\text{out}^{+}=0$ when $%
x^{+}=0$. As the coordinate function $x_\text{out}^{+}(x^{+})$ is a
monotonically increasing function of $x^{+}$, it has the inverse function $%
x^{+}=F_\text{out}\left( x_\text{out}^{+}\right) $, which satisfies $x=F_\text{out}\left(
x_\text{out}^{+}(x)\right) $, i.e.,
\begin{equation}
x^{+}=F_\text{out}\left( x_\text{out}^{+}\right)\quad \Leftrightarrow\quad
x_\text{out}^{+}(x^{+})=\int_{0}^{x^{+}}dy\exp \left( \Theta\! \left(y+\frac{L}{2}%
\right) \right).
\end{equation}%
Similarly, the coordinate function $x_\text{out}^{-}( x^{-}) $ can be
derived from Eq.~(\ref{e3}) to be%
\begin{equation}
x_\text{out}^{-}(x^{-}) =x_\text{out}^{+}(x^{-}-L)
+L=\int_{0}^{x^{-}-L}dy\exp \left( \Theta\!\left(y+\frac{L}{2}%
\right) \right)+L,  \label{e9}
\end{equation}
where Eq.~(\ref{e8}) is used within the right hand side. In the same way as the above discussion, it is possible to introduce
continuous conformally flat coordinates $x_\text{in}^{\pm }$, which connect  Region 
I and  Region II. The coordinate functions $x_\text{in}^{+}(x^{+})$ and $%
x_\text{in}^{-}(x^{-})$ obey
\begin{equation}
x_\text{in}^{+}\!\left(x^{0}+\frac{L}{2}\right)
-x_\text{in}^{-}\!\left(x^{0}-\frac{L}{2}\right) =L,
\label{e11}
\end{equation}%
at the boundary $x^1=x^1_\text{in}=L/2$. As  Region I remains flat,
the metric form is given by $ds^{2}=-dx_\text{in}^{+}\,dx_\text{in}^{-}$  for $x^1_\text{in}>%
L/2$. By changing $L\rightarrow -L$ in Eq.~(\ref{e8}), we obtain the
following relation:
\begin{equation}
x_\text{in}^{+}(x^{+})=\int_{0}^{x^{+}}dy\exp \left( \Theta\! \left(y-\frac{L}{2}%
\right) \right) .
\label{eq:xin}
\end{equation}%
Here, we introduce the inverse function $x^{+}=F_\text{in}\!\left(
x_\text{in}^{+}\right) $, which satisfies $x=F_\text{in}\!\left( x_\text{in}^{+}(x)\right) $ such that
\begin{equation}
x^{+}=F_\text{in}\!\left( x_\text{in}^{+}\right)\quad \Leftrightarrow\quad
x_\text{in}^{+}(x^{+})=\int_{0}^{x^{+}}dy\exp \left( \Theta\! \left(y-\frac{L}{2}%
\right) \right).
\end{equation}%
In Eq.~(\ref{e9}), changing $L\rightarrow -L$ yields 
\begin{equation}
x_\text{in}^{-}\!\left( x^{-}\right) =x_\text{in}^{+}\!\left( x^{-}+L\right)
-L=\int_{0}^{x^{-}+L}dy\exp \left( \Theta\!\left(y-\frac{L}{2}%
\right) \right)-L.
\end{equation}

\subsection{Conformal Symmetry of Expanding Edge Excitations}

In this section, we introduce the coordinate transformation between $x_\text{out}^{+}$
and $x_\text{in}^{+}$. Let $F_\text{in}^{-1}(x)$ denote the inverse function of $%
F_\text{in}(x)$, i.e., $F_\text{in}^{-1}(F_\text{in}(x))=x$. It should be noted that  $F_\text{out}^{-1}(F_\text{out}(x))=x$ holds true. We then define the composite function $%
F(x)=F_\text{in}^{-1}\left( F_\text{out}(x)\right) $, which satisfies $%
x_\text{in}^{+}=F(x_\text{out}^{+})$. From Eq.~(\ref{e8}) and Eq.~\eqref{eq:xin}, the following relation holds:
\begin{equation}
X=\int_{L/2}^{F_\text{out}\left( X\right)+L/2 }dy\exp \left( \Theta\! \left(y%
\right) \right) ,\quad
F(X) = \int_{-L/2}^{F_\text{out}\left(
X\right)-L/2}dy\exp \left( \Theta\!\left(y\right) \right)
.
\label{ee11}
\end{equation}%
By combining two equations in Eq.~(\ref{ee11}), the following relation is derived:
\begin{equation}
F(X) =X-\int_{F_\text{out}\left( X\right) -L/2}^{F_\text{out}\left(
X\right) +L/2}dy \exp \left( \Theta\!\left(y\right) \right)+
\int_{-L/2}^{L/2}dy \exp \left( \Theta\!\left( y\right) \right).
\label{ee110}
\end{equation}%
If the function $\Theta\!\left(v\,t\right) $ is a constant function
independent of $t$, the above equations reduce to $F(X) =X$. However,
if $\Theta\!\left(v\, t\right) $ depends on $t$, then $F(X) $ is a
nontrivial function, which describes the dynamics of the expanding edge region.

If the conformal symmetry (Virasoro symmetry) survives even in the
expansion, the time evolution of the edge excitation is given in Region I and
Region II simply by $\varphi _\text{in}\!\left( x_\text{in}+v\,t_\text{in}\right) $, where $%
\varphi _\text{in}\!\left( x_\text{in}^{+}\right) $ is the initial configuration in
Region I. In Region II, the configuration can be expressed with respect
to $x_\text{out}^{+}$ as $\varphi _\text{in}\!\left( F\!\left( x_\text{out}^{+}\right) \right) $%
. Assuming the conformal symmetry, the configuration in Region III takes the same form as that within Region II:
\begin{equation}
\varphi _\text{out}(x_\text{out}+v\,t_\text{out})=\varphi _\text{out}(x_\text{out}^{+})=\varphi
_\text{in}\!\left( F\!\left( x_\text{out}^{+}\right) \right) .  \label{e12}
\end{equation}%
In future experiments of the expanding edge systems, confirmation of the above
equation directly implies verification of the conformal symmetry
preservation.

\subsection{Measurement of Bogoliubov Coefficients and its Implications}

In this section, we consider the dynamical analysis of excitations in the
expanding edge using the Bogoliubov coefficients in the curved spacetime field
theory. Suppose plane wave mode functions are defined by
\begin{equation}
u_{k}(x)=\frac{e^{ -ikx}}{\sqrt{4\pi k}} ,  \label{e0}
\end{equation}%
where the wave number $k$ has continuous positive values. Then the input wave $%
\varphi _\text{in}(x_\text{in}^\text{+})$ in  Region I can be expanded as
\begin{equation}
\varphi _\text{in}(x_\text{in}^{+})=\int_{0}^{\infty }dk\left(
a_{k}^{(\text{in})}\,u_{k}(x_\text{in}^{+})+a_{k}^{(\text{in})\ast }\,u_{k}^*(x_\text{in}^{+})\right) ,
\end{equation}%
where $a_{k}^{(\text{in})}$ are complex coefficients in this expansion. From Eq.~(%
\ref{e12}), the output wave is then calculated as
\begin{equation}
\varphi _\text{out}(x_\text{out}^{+})=\int_{0}^{\infty }dk'\left(
a_{k'}^{(\text{in})}\,u_{k'}(F\!\left( x_\text{out}^{+}\right)
)+a_{k'}^{(\text{in})\ast }\,u_{k'}^*(F\!\left( x_\text{out}^{+}\right)
)\right) .
\end{equation}%
Note that the output wave form is given by $\varphi_{\text{out},k'} \left(
x_\text{out}^{+}\right)=u_{k'}(F\!\left(
x_\text{out}^{+}\right) )$ if the input wave is the plane wave as in Eq. (\ref{e0}) with $k=k'$.
To extract information about the dynamics in  Region II induced by
the edge expansion, let us expand $u_{k^{\prime }}(F\!\left(
x_\text{out}^{+}\right) )$ with respect to the plane wave mode functions as
\begin{equation}
u_{k^{\prime }}(F\!\left( x_\text{out}^{+}\right) )=\int_{0}^{\infty }dk\left( \alpha (k,k^{\prime })\,
u_{k}\!\left(x_\text{out}^{+}\right)+\beta (k,k^{\prime })\,u_{k}^*\!\left(
x_\text{out}^{+}\right)\right). \label{ee1}
\end{equation}%
The coefficients $\alpha (k,k^{\prime })$ and $\beta (k,k^{\prime })$ are
referred to as Bogoliubov coefficients and satisfy the unitarity conditions
\cite{BD} as follows: 
\begin{subequations}
\begin{align}
\int_{0}^{\infty }dq\left( \alpha (k,q)\,\alpha^* (k^{\prime },q)-\beta
(k,q)\,\beta^* (k^{\prime },q^{\prime })\right) & =\delta \left(
k-k^{\prime }\right) ,  \label{e24} \\
\int_{0}^{\infty }dq\left( \alpha (k,q)\,\beta (k^{\prime },q)-\beta
(k,q)\,\alpha (k^{\prime },q)\right) & =0. \label{e25}
\end{align}%
\end{subequations}
The output wave $\varphi _\text{out}(x_\text{out}^{+})$ can be expanded using the plane wave mode functions in Eq.~(\ref{e0}) as
\begin{equation}
\varphi _\text{out}(x_\text{out}^{+})=\int_{0}^{\infty }dk\left(
b_{k}^{(\text{out})}\,u_{k}(x_\text{out}^{+})+b_{k}^{(\text{out})\ast }\,u_{k}^*(x_\text{out}^{+})\right) ,
\end{equation}%
where $b_{k}^{(\text{out})}$ are the expansion coefficients and computed as follows:
\begin{equation}
b_{k}^{(\text{out})}=\int_{0}^{\infty }dk^{\prime }\left( a_{k^{\prime
}}^{(\text{in})}\,\alpha (k,k^{\prime })+a_{k^{\prime }}^{(\text{in})\ast }\,\beta^*
(k,k^{\prime })\right) .
\end{equation}
The coefficients $\alpha (k,k^{\prime })$ and $\beta
(k,k^{\prime })$ play a crucial role in the quantum regime \cite%
{BD}. By introducing creation operators $\hat{a}_{k}^{(\text{in})\dag }\,,\hat{b}%
_{k}^{(\text{out})\dag }$ and annihilation operators $\hat{a}_{k}^{(\text{in})}, \hat{b}%
_{k}^{(\text{out})}$, satisfying 
\begin{equation}
\left[ \hat{a}_{k}^{(\text{in})},~ \hat{a}_{k^{\prime }}^{(\text{in})\dag }\right] =\hbar\,
\delta( k-k^{\prime }) ,\quad\left[ \hat{b}_{k}^{(\text{out})},~ \hat{b}%
_{k^{\prime }}^{(\text{out})\dag }\right] =\hbar\, \delta( k-k^{\prime }),
\end{equation}%
the field operators $\hat{\varphi}_\text{in}(x_\text{in}^{+}),\hat{\varphi}%
_\text{out}(x_\text{out}^{+})$ are given by
\begin{subequations}
\begin{eqnarray}
\hat{\varphi}_\text{in}(x_\text{in}^{+}) &=&\int_{0}^{\infty }dk\left( \hat{a}%
_{k}^{(\text{in})}\,u_{k}\left( x_\text{in}^{+}\right) +\hat{a}_{k}^{(\text{in})\dag }\,u_{k}^*\left(
x_\text{in}^{+}\right)\right) ,  \label{e20} \\
\hat{\varphi}_\text{out}(x_\text{out}^{+}) &=&\int_{0}^{\infty }dk\left( \hat{b}%
_{k}^{(\text{out})}\,u_{k}(x_\text{out}^{+})+\hat{b}_{k}^{(\text{out})\dagger }\, u_{k}^*(x_\text{out}^{+})\right) .  \label{e21}
\end{eqnarray}%
\end{subequations}
Using $\alpha (k,k^{\prime })$ and $\beta (k,k^{\prime })$, we obtain proof of the following
relations: 
\begin{subequations}
\begin{align}
\hat{b}_{k}^{(\text{out})}& =\int_{0}^{\infty }dk^{\prime }\left( \alpha
(k,k^{\prime })\,\hat{a}_{k^{\prime }}^{(\text{in})}+\beta^* (k,k^{\prime })\,\hat{a}%
_{k^{\prime }}^{(\text{in})\dag }\right) ,  \label{50} \\
\hat{b}_{k}^{(\text{out})\dag }& =\int_{0}^{\infty }dk^{\prime }\left( \alpha^*
(k,k^{\prime })\,\hat{a}_{k^{\prime }}^{(\text{in})\dag }+\beta (k,k^{\prime
})\, \hat{a}_{k^{\prime }}^{(\text{in})}\right) .  \label{51}
\end{align}
\end{subequations}

Note that the input vacuum state $|0_\text{in}\rangle $ is defined by $\hat{a}%
_{k}^{(\text{in})}|0_\text{in}\rangle =0$. Thus, $\beta (k,k^{\prime })$ describes
particle creation in Region III from the input vacuum in Region I. The
expectation values of particle number density $\hat n_{k}^{(\text{out})}$, with wave
number $k$, are computed as 
\begin{equation}
\left\langle \hat n_{k}^{(\text{out})}\right\rangle =\frac{1}{\hbar }\,\langle 0_\text{in}|\,\hat{%
b}_{k}^{(\text{out})\dag }\,\hat{b}_{k}^{(\text{out})}\,|0_\text{in}\rangle =\int_{0}^{\infty
}dk^{\prime }\left\vert \beta (k,k^{\prime })\right\vert ^{2}.
\end{equation}%
The total number of particles is evaluated as 
\begin{equation*}
\left\langle \hat N^{(\text{out})}\right\rangle =\int_{0}^{\infty }dk\left\langle
\hat n_{k}^{(\text{out})}\right\rangle.
\end{equation*}%
This implies that quantum particle creation in the expanding analog universes 
can be predicted only from $\beta (k,k^{\prime })$, which can be measured
in the classical regime of the system. Thus, experimentally determining $\beta
(k,k^{\prime })$ is crucial.

Here it may be useful to comment about a relation between the conformal symmetry and the coefficient $\beta (k,k^{\prime })$. If a conformally flat coordinate system covers the entire spacetime region (Region I + Region II + Region III) in the experiments and remains the flat metric in Region I and Region III, $x_\text{out}^+$ is equal to $x_\text{in}^+$ up to a factor and a constant. Thus the conformal symmetry always results in $\beta (k,k^{\prime })=0$. It should be stressed that this is not the case in our situation since we do not have such a global conformally flat coordinate system. Due to this fact, the function $u_{k^{\prime }}(F\left( x_\text{out}^{+}\right))$ becomes a nontrivial function with a transient behavior in a region of $ x_\text{out}^{+}$, and includes negative frequency out modes. This implies that  $\beta (k,k^{\prime })\neq 0$ in general even if the conformal symmetry is preserved. 

As Region I and  Region III remain flat during the expansion of  Region 
II, the excitations in the regions are always described by the free fields
in Eq.~(\ref{e20}) and Eq.~(\ref{e21}). The coefficients $\alpha (k,k^{\prime
})$ and $\beta (k,k^{\prime })$ can be estimated using the experimental
data in  Region III, even if the conformal symmetry is broken in  Region II. Thus, the question of whether the conformal symmetry survives can be
answered by the experimental data analysis of the wave form $\varphi_\text{out}(x)$ in Eq.~(\ref{e12}), $\alpha (k,k^{\prime })$, and $\beta (k,k^{\prime })$ in Eq.~(\ref{ee1}). 
Let us consider a deviation $\delta \varphi_{\text{out},k'}(x)$ from the predicted function form $\varphi_{\text{out},k'}(x)$. Then the following relation holds for the deviations of $\alpha (k,k^{\prime })$ and $\beta (k,k^{\prime })$:
\begin{equation}
\delta \varphi_{\text{out},k'}(x_\text{out}^{+})=\int_{0}^{\infty }dk\left( \delta \alpha (k,k^{\prime })\,
u_{k}\!\left(x_\text{out}^{+}\right)+\delta \beta (k,k^{\prime })\,u_{k}^*\!\left(
x_\text{out}^{+}\right)\right). \label{eeee1}
\end{equation}
From the above equation, the following relations are directly computed via the Fourier transformation:
\begin{align}
&\delta \alpha (k,k^{\prime })=\sqrt{\frac{k}{\pi}} \int^{\infty}_{-\infty} dx^+\, \delta \varphi_{\text{out},k'} (x^{+}) \exp\left(ik x^+ \right),\label{r1} \\
&\delta \beta (k,k^{\prime })=\sqrt{\frac{k}{\pi}} \int^{\infty}_{-\infty} dx^+\, \delta \varphi_{\text{out},k'} (x^{+}) \exp\left(-ik x^+ \right).\label{r2}
\end{align}
By substituting observed deviation $\delta \varphi_{\text{out},k'}$ into Eq. (\ref{r1}) and Eq. (\ref{r2}), 
we are able to quantify the conformal symmetry breaking for every wavelength $k'$ of the input wave and wavelength $k$ of the output plane wave. It is possible that the conformal symmetry breaking may occur only for some regions of $(k,k')$. In such a case, Eq. (\ref{r1}) and Eq. (\ref{r2}) provide the information of detailed structure of the symmetry breaking.

If the future experiments will confirm breakdown of the conformal symmetry, then we will be able to consider various kinds of 
extended effective field theories in Region II to compute $\alpha
(k,k^{\prime })$ and $\beta (k,k^{\prime })$, and compare the predictions with the
experimental results.  To construct simple effective theories describing the possible breakdown of the conformal symmetry, we propose to assume the general covariance, i.e., symmetry under general coordinates transformations as a   working hypothesis for the dynamics. Then the equation of motion takes a general form as follows:
\begin{equation}
\left( \nabla ^{2}-\left(\frac{m(t)v}{\hbar}\right)^2-\xi R(t)\right) \varphi (t,x)-U^{\prime }(
\varphi,t) =J(t,x).  \label{e22}
\end{equation}%
Here $\nabla ^{2}$ is the covariant Laplacian operator in  Region II, $m(t)$
is the time-dependent effective mass induced by the expansion. The scalar
curvature of the FLRW spacetime in  Region II is denoted by $R(t)$, and $\xi $ is a real parameter controlling the curvature interaction. The terms with $m(t)$ and $\xi R(t)$ induce different velocity of the edge excitation in  Region II for each wave number $k$, and break the conformal symmetry. The term $U^{\prime }( \varphi, t) $ 
represents a time-dependent nonlinear interaction of the field $\varphi (t,x)$, which may also break the conformal symmetry. The term $J(t,x)$ is a possible source. When the source term $J(t,x)$ exists, an additional wave $\varphi_J (t,x)$ is classical-mechanically generated by the source, even in the zero input wave case. Note that the source $J(t,x)$ also breaks the conformal symmetry, but it remains harmless when the nonlinear interaction does not exist,  or when the amplitude of the input wave is so small that the nonlinear interaction can be neglected. In such cases, it turns out that the source term can be eliminated by subtracting $\varphi_J (t,x)$  from $\varphi (t,x)$. The dynamics information about the edge excitations in the Region II  is imprinted onto $\alpha (k,k^{\prime })$ and $\beta (k,k^{\prime })$, which enables us to check
the validity of Eq.~(\ref{e22}).

Increasing the amplitude of the initial wave $\varphi _\text{in}(x_\text{in}^{+})$
in  Region I, the experiments are able to  determine whether nonlinear interactions like $%
U^{\prime }( \varphi,t) $ are  generated in  Region II. If
the amplitude of $\varphi _\text{out}(x_\text{out}^{+})$ does not increase linearly as
the amplitude of $\varphi _\text{in}(x_\text{in}^{+})$, the nonlinear interaction exists.
Furthermore, it should be emphasized that an experimental check of the edge unitarity
conditions of $\alpha (k,k^{\prime })$ and $\beta (k,k^{\prime })$ in Eq.~(\ref%
{e24}) and Eq.~(\ref{e25}) may detect the existence of bulk-edge interactions
in  Region II. If the edge excitations interact with bulk excitations like
magneto-roton, the edge dynamics are coupled with the bulk dynamics and cannot be determined only by the edge information. Specifically, the relations in Eq.~(%
\ref{e24}) and Eq.~(\ref{e25}) may be extended to the following relations: 
\begin{align*}
&\int_{0}^{\infty }\!\!dq\left( \alpha (k,q)\,\alpha^* (k^{\prime },q)-\beta (k,q)\,\beta^*(k^{\prime },q)\right)
+\int\!\! d^2q\left( \tilde{\alpha}(k,\vec{q})\,\tilde{\alpha}^*(k^{\prime },%
\vec{q})-\tilde{\beta}(k,\vec{q})\,\tilde{\beta}^*(k^{\prime },\vec{q}%
)\right) 
=\delta \left( k-k^{\prime }\right) ,\\
&\int_{0}^{\infty }\!\!dq\left( \alpha (k,q)\,\beta (k^{\prime },q)-\beta
(k,q)\,\alpha (k^{\prime },q)\right) 
+\int \!\!d^{2}q\left( \tilde{\alpha}(k,\vec{q})\,\tilde{\beta}(k^{\prime },\vec{%
q})-\tilde{\beta}(k,\vec{q})\,\tilde{\alpha}(k^{\prime },\vec{q})\right)
=0,
\end{align*}%
where $\tilde{\alpha}(k,\vec{q})$ and $\tilde{\beta}(k^{\prime },\vec{q})$
are contributions to the Bogoliubov coefficients induced by the bulk mode
functions. This implies that Eq.~(\ref{e24}) and Eq.~(\ref{e25}) are not satisfied
if the bulk-edge interactions appear in  Region II. Consequently, the experiments are able to provide upper bounds of coupling constants of  bulk-edge interactions in various situations.
More detailed analyses in this section will be reported in forthcoming papers.

\section{Expanding Edges as Universe Simulators in 2D Dilaton Gravity}

In this section, we show that the expanding edges of QH systems are regarded
as universe simulators of 2D dilaton gravity models. Some useful formula for
2D gravity are given in the Appendix. Let us suppose the following action for a
real scalar field $\Phi $,\ which is called the dilaton field, and the metric field $%
g_{\alpha \beta }$:
\begin{equation}
S=\int d^{2}x\sqrt{-g}\left( \Phi R-4\lambda^{2}\,V(\Phi)
\right) ,  \label{13}
\end{equation}%
where $\lambda$ is a positive constant and $V(\Phi) $ is the potential term of $\Phi $. As shown later, the determination of the time schedule of edge expansion corresponds to the determination of a form of $V(\Phi) $. Thus, the field $\varphi$ in Eq.~(\ref{e22}) can be interpreted as a matter field in a curved spacetime background of 2D dilation gravity with the potential term $V(\Phi)$. 
The metric
equation $\delta S/\delta g^{\alpha \beta }(x)=0$ derived from Eq.~(%
\ref{13}) is written as%
\begin{equation}
\left( g_{\alpha \beta }\nabla ^{2}-\nabla _{\alpha }\nabla _{\beta }\right)
\Phi +2\lambda^{2}\,V(\Phi)\, g_{\alpha \beta }=0.  \label{15}
\end{equation}%
Here we have used a property of the 2D gravity, such that 
$R_{\alpha\beta}-R\,g_{\alpha\beta}/2=0$.
The field equation $\delta S/\delta \Phi (x)=0$ reads 
\begin{equation}
R=4\lambda ^{2}\,V^{\prime }(\Phi) ,  \label{16}
\end{equation}%
where $V^{\prime }( \Phi) $ is the derivative function of $V(\Phi) $ with respect to $\Phi $. As already mentioned in Sec. II, any 2D spacetime is able to be described, at least locally,
by the conformally flat metric as $
ds^{2}=-\exp \left( 2\Theta \left( x^{+},x^{-}\right) \right) dx^{+}dx^{-}$.
In this coordinate system, Eq.~(\ref{15}) is simplified to the following three
equations:
\begin{subequations}
\begin{align}
&\partial _{+}\partial _{-}\Phi -\lambda ^{2}\,V(\Phi)\, e^{2\Theta
}=0,  \label{17} \\
&\partial _{+}^{2}\Phi -2\partial _{+}\Theta\, \partial _{+}\Phi =0,  \label{18}\\
&\partial _{-}^{2}\Phi -2\partial _{-}\Theta\, \partial _{-}\Phi =0.  \label{19}
\end{align}%
\end{subequations}
Similarly Eq.~(\ref{16}) is reduced to
\begin{equation}
2\partial _{+}\partial _{-}\Theta -\lambda ^{2}\,V^{\prime }( \Phi)\, e^{2\Theta }=0.  \label{20}
\end{equation}
To describe FLRW universes, let us consider metric forms
such that $\Theta $ does not have any $x$ dependence: 
\begin{equation*}
ds^{2}=-e^{2\Theta ((x^{+}+x^{-})/2
)}\,dx^{+}dx^{-}=e^{2\Theta (vt)}\left( -v^{2}dt^{2}+dx^{2}\right) .
\end{equation*}%
The conformal factor $e^{2\Theta (vt)}$ directly corresponds to the
conformal factor in Eq.~(\ref{e2}) of the expanding-edge realistic
experiments. In this case, Eq.~(\ref{17}) and Eq.~(\ref{20}) become
\begin{subequations}
\begin{align}
&\frac{d^{2}\Phi }{dt^{2}}-4v^{2}\lambda^{2}\,V(\Phi(t))
\,e^{2\Theta (vt)}=0, \label{21}\\
&\frac{d^{2}\Theta }{dt^{2}}-2v^{2}\lambda^{2}\,V'(\Phi
(t))\, e^{2\Theta (vt)}=0.  \label{23}
\end{align}%
\end{subequations}
Equations~(\ref{18}) and (\ref{19}) result in the same equation as
\begin{equation}
\frac{d^{2}\Phi }{dt^{2}}-2\frac{d\Theta }{dt}\frac{d\Phi }{dt}=0.
\label{22}
\end{equation}%
Let us divide both sides by $(d\Phi/dt)$ in Eq.~(\ref{22}), yielding
$
d/dt\left( \ln( d\Phi/dt)-2\Theta (v\,t)\right) =0
$.
By integrating the aforementioned equation with respect to $t$, we obtain
$d\Phi/dt=A\,e^{2\Theta (vt)}$
where $A$ is a positive integration constant. The constant $A$ can be replaced by $ \lambda\, v$ using a transformation such that 
$
\Theta (vt) \rightarrow \Theta (vt)-1/2\ln A+1/2\ln(\lambda v)$.
Thereby, the equation can be written as
\begin{equation}
\frac{d\Phi }{dt}=  \lambda v \,e^{2\Theta (vt)}.  \label{25}
\end{equation}%
By integrating Eq.~(\ref{25}), we obtain the following equation:
\begin{equation}
\Phi (t)= \lambda v \int_{0}^t dt'\,e^{2\Theta (vt^{\prime })} +\Phi_{0},
\label{eq:phi-evo}
\end{equation}
where $\Phi_{0}$ is a constant. In the expanding-edge experiment, the factor 
$e^{2\Theta (vt)}$ is fixed as a function of $t$. This determines the
monotonically increasing function $\Phi (t)$ in Eq.~\eqref{eq:phi-evo}.  Subtracting
Eq.~(\ref{21}) from Eq.~(\ref{22}) yields 
\begin{equation*}
2\frac{d\Theta }{dt}\frac{d\Phi }{dt}=4v^{2}\lambda ^{2}\,V(\Phi
(t))\, e^{2\Theta (v\,t)}.
\end{equation*}%
By substituting Eq.~(\ref{25}) into the above equation, the potential as
a function of time $t$ can be determined by 
\begin{equation}
V(\Phi (t)) =\frac{1}{2v \lambda} \frac{d\Theta }{dt}(v\,t).
\label{28}
\end{equation}%
Here we define
the inverse function of $\Phi (t)$ as $t\left( \Phi \right) $. 
Since $\Phi (t\left( \Phi \right) )=\Phi $ holds, the potential term $%
V(\Phi) $ is fixed by the function $\Theta (vt)$ as
\begin{equation}
V(\Phi) =\frac{1}{2v \lambda} \frac{d\Theta }{dt}%
(v\,t\left( \Phi \right) ).  \label{30}
\end{equation}%
It is worth highlighting that various function forms of $\Theta (vt)$ can be
realized by changing the time schedule of the edge expansion in the experiments.
This implies that the expanding edge experiments provide universe simulators
of the dilaton gravity models.

When we consider $V(\Phi) =-\Phi $, the model becomes the
Jackiw--Teitelboim (JT) model \cite{T, J}. The equation of motion
possesses a solution of anti-de Sitter (AdS) \ spacetime with negative
constant curvature $R=-4\lambda ^{2}$. The edge expansion experiment is
capable of simulating the JT model by taking the conformal factor as
\begin{equation}
e^{\Theta (vt)}=\frac{1}{\cosh \left(H vt\right) },
\end{equation}%
where we introduced  $H=\sqrt{2}\,\lambda$; $H^{-1}$ corresponds to the AdS curvature radius.
The conformal factor yields the following FLRW\ metric of AdS spacetime with proper time $\tau$:
\begin{eqnarray}
ds^{2}=-v^{2}d\tau ^{2}+\cos ^{2}\left(H v\tau \right) dx^{2}.
\label{AdeSitter}
\end{eqnarray}%
From Eq.~(\ref{eq:phi-evo}), the time evolution of the dilaton field is given by 
\begin{equation}
\Phi (t)=\frac{Hv}{\sqrt{2}}\int_{0}^{t}%
\frac{dt^{\prime }}{\cosh ^{2}\left(H vt^{\prime }\right) }+\Phi_0=%
\frac{1}{\sqrt{2}}\tanh \left(Hvt\right)+\Phi_0 .
\label{32}
\end{equation}%

Next, let us consider the dSJT model, which is constructed by $H\rightarrow iH$ in the JT model with the negative cosmological constant $-\lambda ^{2}$. The corresponding potential term in Eq.~\eqref{13} is
given by $V(\Phi) =\Phi $. This model reproduces the de Sitter
spacetime with a positive constant curvature as
$R =4\lambda ^{2}$. 
The conformal factor is fixed as%
\begin{equation}
e^{\Theta \left( vt\right) }=\frac{1}{\cos \left(H vt\right) }.
\label{eq:conf-DS}
\end{equation}%
In this case, $H^{-1}$ corresponds to the Hubble radius of de Sitter spacetime.
Since the FLRW metric is computed using proper time $\tau $ as 
\begin{equation}
ds^{2}=-v^{2}d\tau ^{2}+\cosh ^{2}\left( H v\tau \right)
dx^{2},
\label{eq:DS}
\end{equation}%
the analog universe experiences an inflationary expansion when $\tau>0$. From Eq.~\eqref{eq:phi-evo}, the time evolution of
the dilaton field is derived as follows:
\begin{equation*}
\Phi (t)=\frac{1}{\sqrt{2}}\tan \left( H vt\right)+\Phi_0 .
\end{equation*}

We demonstrate the effect of the expanding (shrinking) region 
on the edge mode's waveform. The mode function in  Region III
is related to the mode function in  Region I by Eq.~(\ref{e12}), 
which is explicitly obtained by finding the function $F(X)$. 
The function $F(X)$ is explicitly given by Eq.~(\ref{FXdeSitter}) 
when the boundary of  Region II is the de Sitter spacetime (\ref{eq:DS}),
while $F(X)$ for the anti-de Sitter spacetime is given by the analytic continuation. The derivation of $F(X)$ for the de Sitter case is presented in 
the next section. From Eq.~(\ref{e12}), the form of edge
excitation $\varphi _\text{out}(x_\text{out}^{+})$ in Region III is predicted,
assuming the conformal symmetry in Region II.
\begin{figure}[H]
  \centering
  \begin{minipage}[b]{0.45\linewidth}
  \includegraphics[width=1\linewidth]{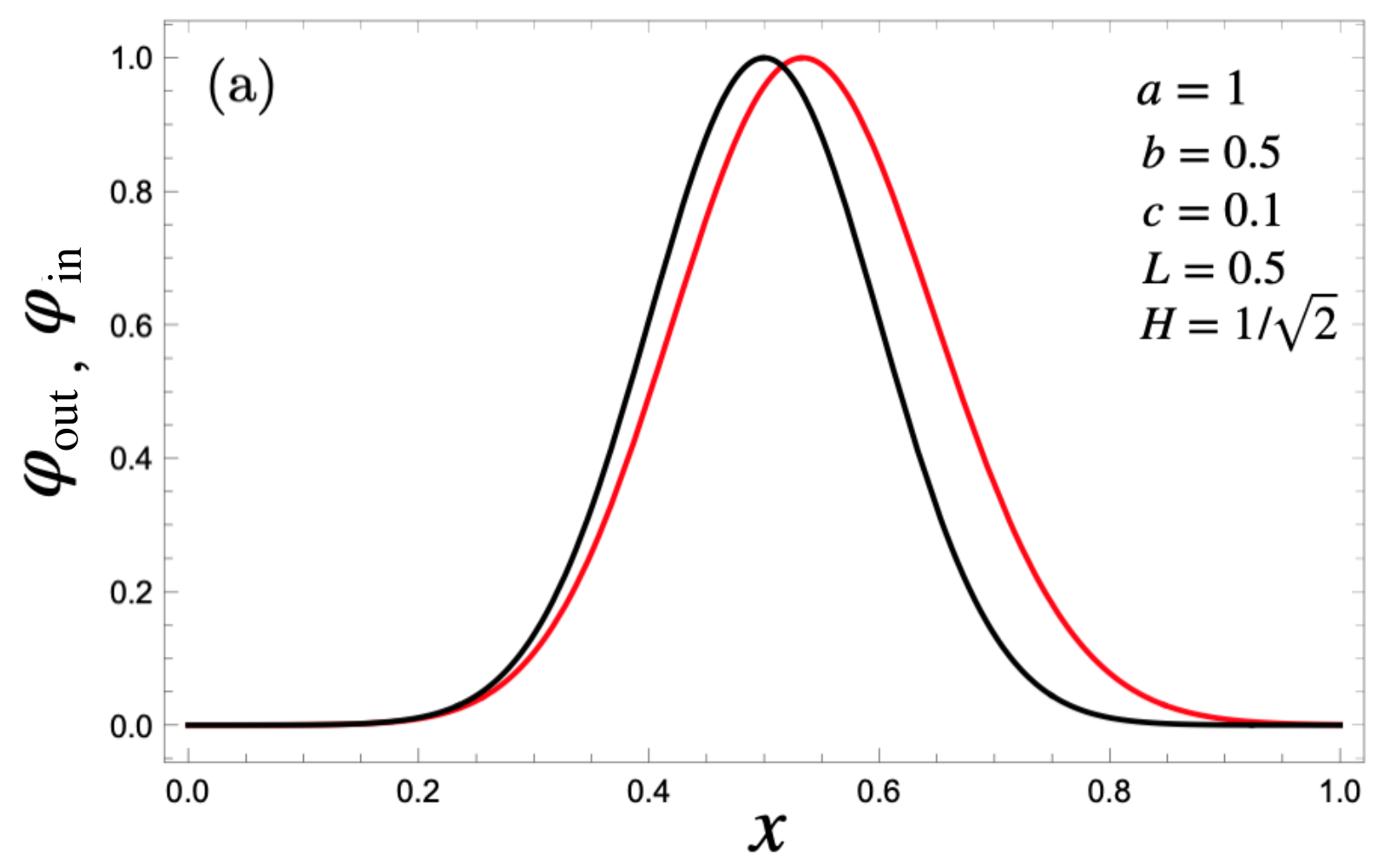}
  \end{minipage}
  \hspace{1cm}
  \begin{minipage}[b]{0.45\linewidth}
  \includegraphics[width=1\linewidth]{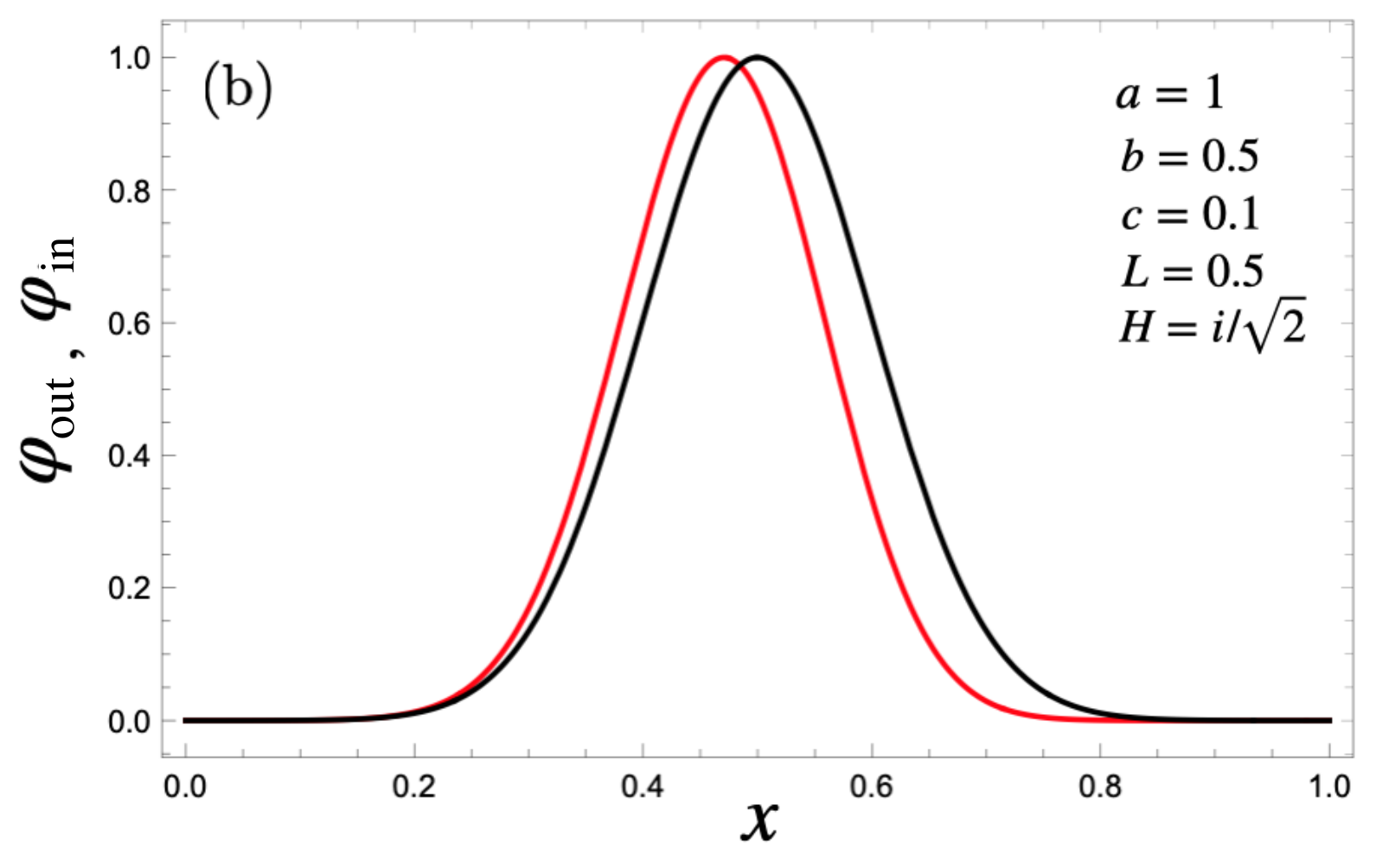}
  \end{minipage}
  \caption{Behavior of $\varphi _{\text{in}}$ (black curves) and $\varphi _{\text{out}}$ (red curves). (a) de Sitter ({$H=1/\sqrt{2}$}) (b) AdS spacetime ({$H=i/\sqrt{2}$}) cases. All parameters, i.e. $a$, $b$, $c$, $L$, are identical except for $H$.
  Here $x$ denotes $x_{\rm in}^+$ 
  for $\varphi_{\rm in}$ and $x_{\rm out}^+$ 
  for $\varphi_{\rm out}$.}
  \label{fig:in-out wave}
\end{figure}
\noindent
Figures~\ref{fig:in-out wave} (a) and (b) show the behavior of the input wave $\varphi _{\text{in}}$ (black curves) and output wave $\varphi _{\text{out}}$ (red curves) in de Sitter and AdS spacetime.
Here, the input wave is assumed to be the following Gaussian form:
\begin{align}
\varphi_{\text {{in}}}(x_{\rm in}^+)=a\exp \left(-\frac{(x_{\rm in}^+-b)^{2}}{2 c^{2}}\right),
\nonumber
\end{align}
where $a, b$, and $c$ are arbitrary parameters.
Note that the parameter $H$ in de Sitter spacetime (\ref{eq:DS}) is replaced by $iH$ in AdS spacetime (\ref{AdeSitter}). 
In our case, de Sitter spacetime is expanding, so the output wave is also spreading, but, in AdS spacetime, it is shrinking, so the output wave is narrowing.
In both cases, the longer the input wave stays in Region II, the more it will be affected by the expansion or shrinkage, and the output wave will behave as if it is expanding or shrinking.

When  Region II is the expanding de Sitter spacetime (\ref{eq:DS}), i.e., inflationary universe, 
there appear the wave modes whose wavelength stretched
infinitely. Namely, the cosmological horizon appears.   
This predicts the Hawking radiation in the edge modes on the
quantum Hall systems, which we will discuss in the next section.

%
\section{Edge mode's Hawking radiation}
\begin{figure}[H]
\centering
\includegraphics[width=0.5\linewidth]{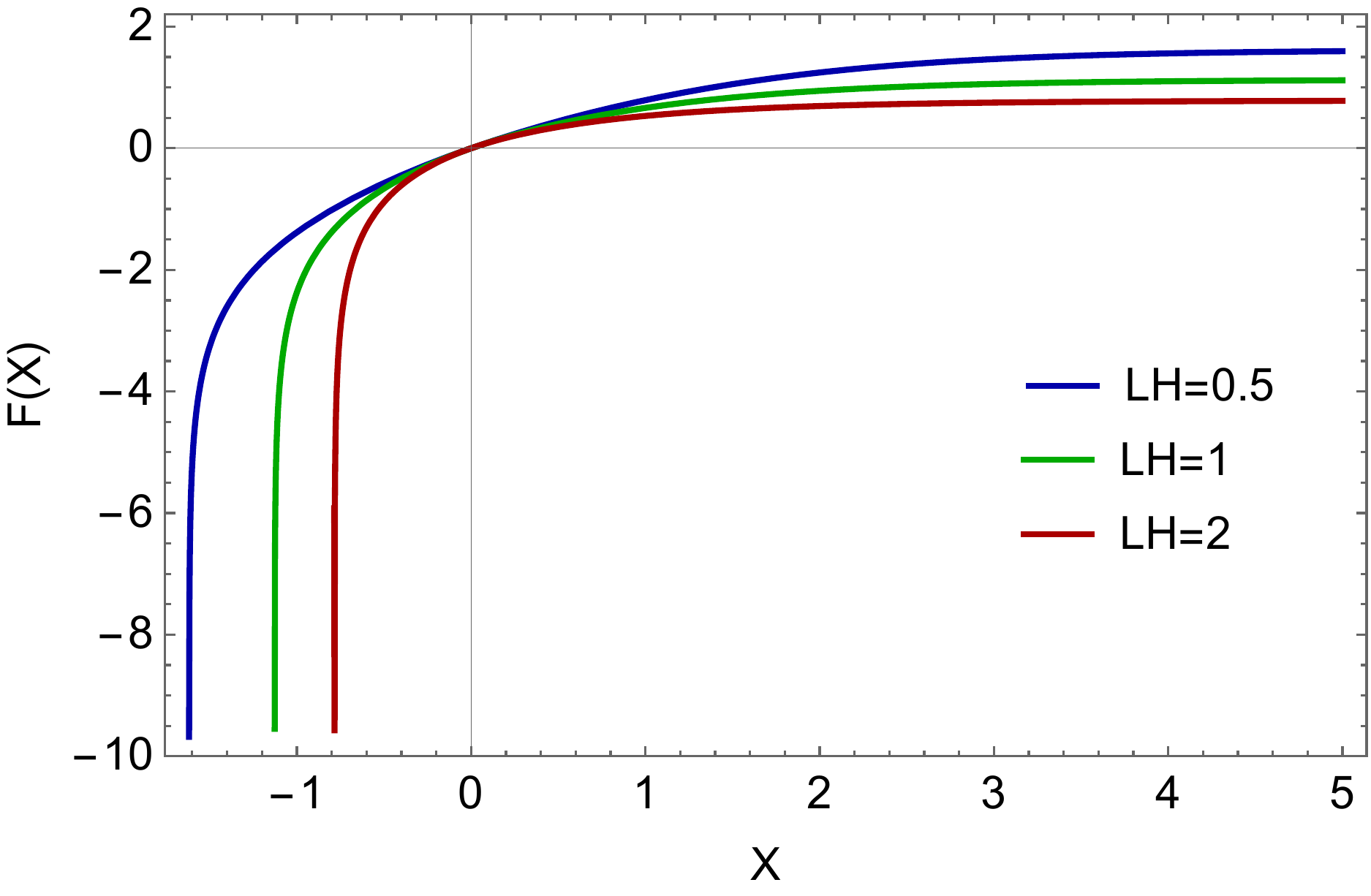}
\caption{Behavior of the function $F(X)$ which relates $x_\text{in}^+$ and $x_\text{out}^+$ as $x_\text{in}^+=F(x_\text{out}^+)$.} The domain and the range of $F(X)$ become semi-indefinite, which indicates existence of horizons in the present spacetime.
\label{fig:F}
\end{figure}

\begin{figure}[H]
\centering
\includegraphics[width=0.8\linewidth]{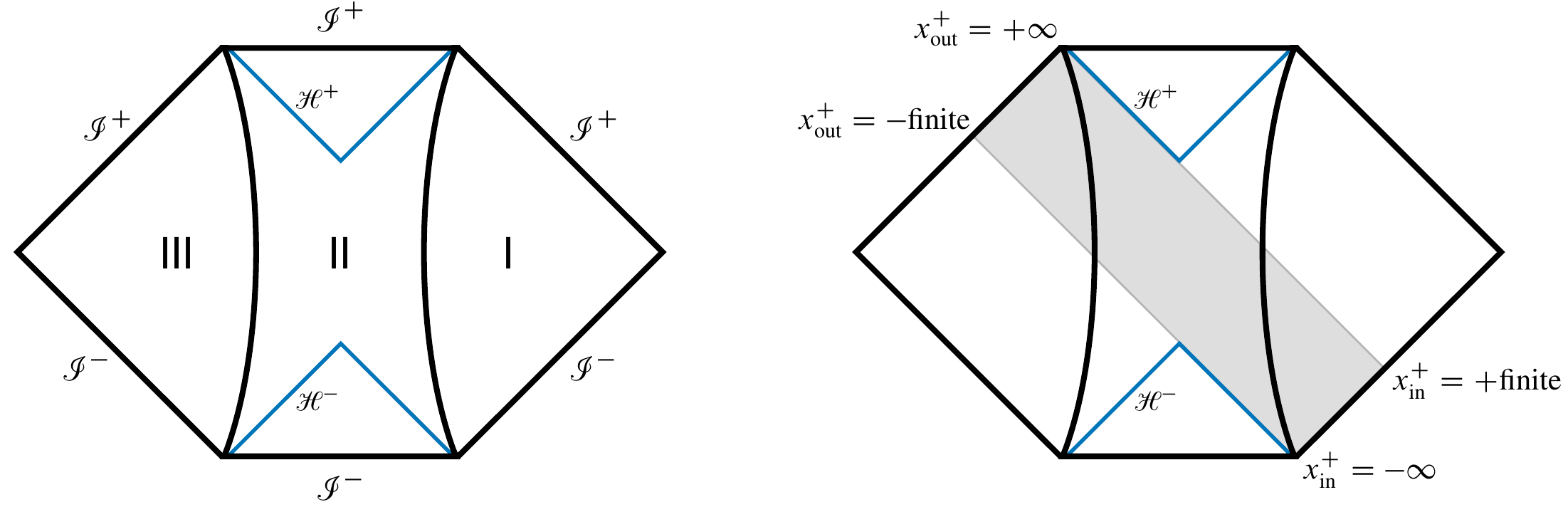}
\caption{Penrose diagram for the QH system with an expanding ($t>0$) and a contracting ($t<0$) edge Region II, which is assumed to be a part of the de Sitter spacetime with the global chart. Owing to its global structure, this spacetime possesses a future horizon $\mathscr{H}^+$ and a past horizon $\mathscr{H}^-$.
The shaded region of the right panel shows the region 
that the edge modes in $\mathscr{I}^-$ in Region I move to reach
$\mathscr{I}^+$ in Region III. }
\label{fig:penrose1}
\end{figure}

\begin{figure}[H]
\centering
\includegraphics[width=0.9\linewidth]{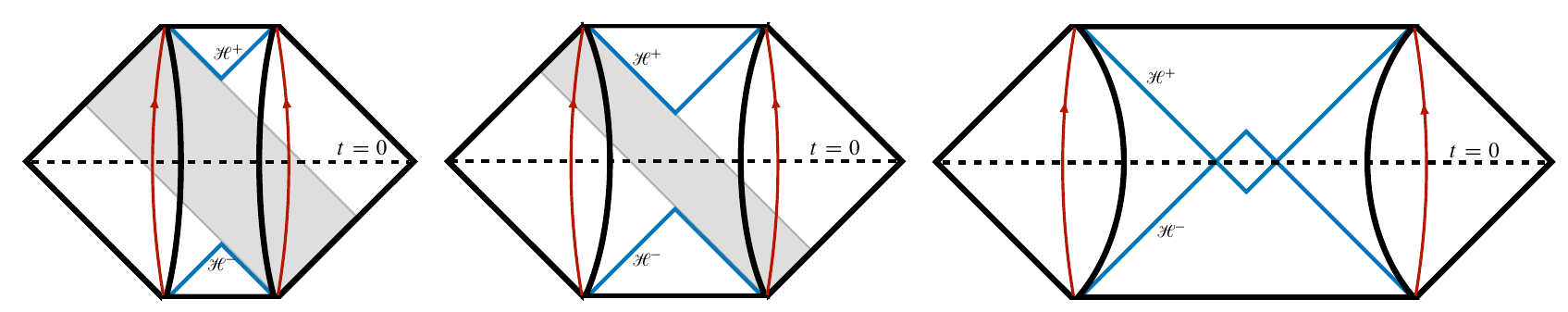}
\caption{$L$ dependence of the global structure of spacetime. Red lines represent world lines for fixed spatial points in Region I and Region III. From left panel to right panel, $0<LH<\pi/4, \pi/4<LH<\pi/2, \pi/2<LH$. For $\pi/2<LH$, signals emitted from the in-region I cannot reach the out-region III.}
\label{fig:penrose2}
\end{figure}

\begin{figure}[H]
\centering
\includegraphics[width=0.5\linewidth]{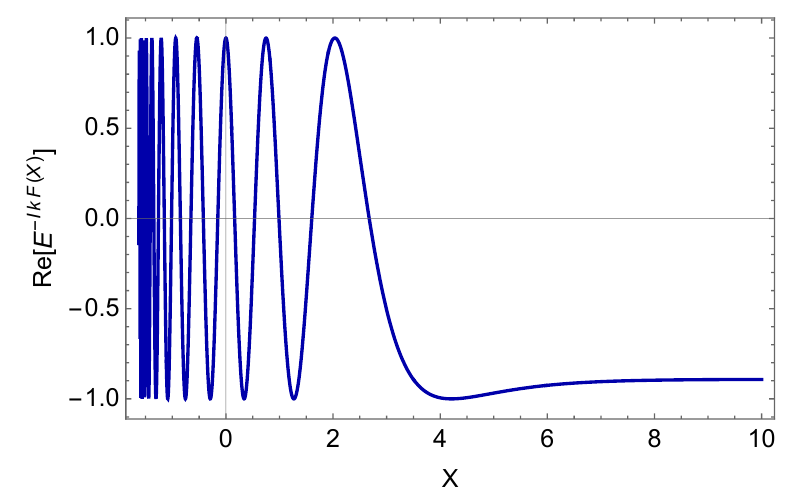}
\caption{The wave form received at a detection point in Region III ($LH=0.5$). $X$ corresponds to $x_\text{out}^+$. The input signal is a plane wave $\varphi_\text{in}=e^{-ikX}$ with $k=10$. For $X\rightarrow X_*\approx -1.6$, the input wave is infinitely blue-shifted owing to the past de Sitter horizon $\mathscr{H}^-$ in Region II, and for $X\rightarrow+\infty$, the input wave is infinitely red-shifted owing to the future de Sitter horizon $\mathscr{H}^+$ in Region II.}
\label{fig:wave-form}
\end{figure}

In this section, we will discuss the Hawking radiation when 
 Region II is an inflationary universe.
We first evaluate $F_\text{out}\left( X\right) $ in (\ref{ee11}) for the inflationary universe  with the conformal factor \eqref{eq:conf-DS}, which covers a whole (1+1)-dimensional de Sitter spacetime. From Eq.~(\ref{ee11}), $F_\text{out}\left( X\right) $ obeys the following equation: 
\begin{equation*}
X=\frac{1}{2H}\ln \left[ \frac{1+\sin \left( H
\left(F_\text{out}\left( X\right) +L/2\right) \right) }{1-\sin \left( 
H\left(F_\text{out}\left( X\right) +L/2\right) \right) }%
\times \frac{1-\sin \left( H L/2\right) }{1+\sin \left( 
H L/2\right) }\right] .
\end{equation*}%
Thus, the functions $F_\text{out}(X)$ and $F(X)$ are computed as
\begin{eqnarray}
&&F_\text{out}\left( X\right) =-\frac{L}{2}+\frac{1}{H }\arcsin \left[
\frac{\left( 1+\sin \left( H L/2\right) \right) e^{2%
H X}-\left( 1-\sin \left( H L/2\right)
\right) }{\left( 1+\sin \left( HL/2\right) \right) e^{2%
H X}+\left( 1-\sin \left( H L/2\right)
\right) }\right], \\
&&F(X)=\frac{1}{2H}\ln \left[ \frac{1+\sin \left( H
\left(F_\text{out}\left( X\right) -L/2\right) \right) }{1-\sin \left( 
H\left(F_\text{out}\left( X\right) -L/2\right) \right) }
\times \frac{1+
\sin \left( H L/2\right) }{1-\sin \left( 
H L/2\right) }\right].
\label{FXdeSitter}
\end{eqnarray}%
The function $F(X)$ behaves as is shown in Fig.~\ref{fig:F}. For a negative value of $X=X_*<0$, $F(X_*)=-\infty$ and for $X\rightarrow+\infty$, $F\rightarrow\text{const.}$. The domain and the range of $F(X)$ are semi-infinite, and this behavior indicates existence of horizons in the present spacetime. About $X=0$, $F(X)\approx X$ which corresponds to behaviour of flat spacetime. Indeed, in the limit of $L\rightarrow 0$, a whole spacetime region becomes flat and $F(X)=X$ is recovered.
\noindent
Figure~\ref{fig:penrose1} shows Penrose diagrams of the analog spacetime for the QH system with an expanding and contracting edge. 
Regions I and III are flat spacetimes, and  Region II is  a part of the de Sitter spacetime representing the expanding edge, whose metric is described by \eqref{eq:DS} with the global chart. This form of de Sitter metric covers  a whole region of the de Sitter spacetime, and represents a contracting universe for $t<0$ and an expanding universe for $t>0$. Three spacetime Regions I, II, III are joined along  spatial points $x_\text{in}=L/2$ and $x_\text{out}=-L/2$. Owing to its global structure, only a part of data prepared on $\mathscr{I}^-$ of  Region I can reach $\mathscr{I}^+$ of  Region III, which implies existence of the event horizon in this spacetime ($\mathscr{H}^+$ in Fig.~\ref{fig:penrose1}).
\noindent
Reachability of signals from the in-region I to the out-region III depends on the size $L$ of the expanding edge Region II (Fig.~\ref{fig:penrose2}); for $\pi/2<LH$, signals from  Region I cannot reach  Region III. For $LH<\pi/4$, signals prepared at $t=0$ in Region I reach Region III at $t>0$.
Figure \ref{fig:wave-form} shows the wave form $\varphi_\text{out}(X)=\varphi_\text{in}(F(X))$ with $\varphi_\text{in}(X)=e^{-ikX}$. Owing to the existence of the future horizon $\mathscr{H}^+$, the wave is stretched and freezes out as $X\rightarrow\infty$. This behavior of wave is same as that for black hole formation via gravitational collapse. Thus, we expect the emission of Hawking radiation from vicinity of the future horizon $\mathscr{H}^+$ if the scalar field is quantized  and the vacuum condition  for the in-vacuum state is imposed at $\mathscr{I}^-$ in Region I, that corresponds to the Unruh vacuum state in the standard scenario of Hawking radiation via gravitational collapse~\cite{hawking,Hawking1975,BD}. We do not discuss detail of quantum effect in this paper (we will discuss this aspect in our forthcoming paper~\cite{HNSYY2022}), but just investigate the Bogoliubov coefficient which can be obtained from the relation between classical wave modes $\varphi_\text{in}$ and $\varphi_\text{out}$.
\noindent
For the input plane wave $\varphi_\text{in}(X)=e^{-ikX}$, from Eq.~\eqref{ee1}, the Bogoliubov coefficients are obtained by the following Fourier transformation
\begin{equation}
    \alpha(k,k')=\sqrt{\frac{k}{k'}}\int_{-\infty}^\infty dX\,e^{-ik'F(X)}e^{ikX},\quad\beta(k,k')=\sqrt{\frac{k}{k'}}\int_{-\infty}^\infty dX\,e^{-ik'F(X)}e^{-ikX}.
\end{equation}
To extract information of late time particle creations in Region II,  we consider the asymptotic behavior of the function $F(X)$ for $HX\gg 1$:
\begin{equation}
    F(X)\approx -c_0-c_1e^{-HX},
\end{equation}
with
\begin{equation}
    c_0=\frac{2}{H}\ln(\tan(HL/2)),\quad c_1=\frac{4}{H}\sqrt{\frac{1-\sin(HL/2)}{1+\sin(HL/2)}}.
\end{equation}
Using this asymptotic form of $F(X)$, we obtain the Bogoliubov coefficient as
\begin{align}
    \beta(k,k')&\approx\sqrt{\frac{k}{k'}}\,\frac{e^{ik'c_0}}{H}\int_0^\infty dy\,y^{-ik/H-1}\,e^{ik'c_1y} \notag\\
    &=\sqrt{\frac{k}{k'}}\,\frac{e^{ik'c_0}}{H}(-ik'c_1)^{-ik/H}\,\Gamma\left(\frac{ik}{H}\right).
\end{align}
Therefore
\begin{equation}
    |\beta(k,k')|^2\approx\frac{2\pi}{Hk'}\,\frac{1}{\exp\left(2\pi k/H\right)-1},
    \label{eq:beta2}
\end{equation}
and it shows the Planckian distribution with a temperature
\begin{equation}
    T_\text{H}=\frac{H}{2\pi}.
\end{equation}
This temperature coincides with the Gibbons-Hawking temperature in the de Sitter spacetime. Equation \eqref{eq:beta2} is a signal of the classical counterpart of  Hawking radiation from the future horizon in Region II. Although  Region II corresponds to an expanding de Sitter universe, owing to our setup of experiment, the global structure of spacetime resembles the situation of black hole formations by  gravitational collapse, and it is possible to detect Hawking radiation from the analog black hole. The temperature of the emitted  Hawking radiation is the same as that of the de Sitter horizon. More detailed analysis on Hawking radiation from the expanding edge region will be presented in our forthcoming paper~\cite{HNSYY2022}.

\section{Summary}
In this paper, we formulated a general theory for the analysis of future
experiments of the expanding edges of QH systems. The dynamics of edge expansion
is described by the wave form $\varphi_\text{out}(x_\text{out}^{+})$ of edge
excitation measured in the Region III and its Bogoliubov coefficients $\alpha
(k,k^{\prime })$ and $\beta (k,k^{\prime })$. Based on Eq.~(\ref{e12}), the
experiments determine whether the conformal symmetry survives in 
Region II. By increasing the amplitude of the initial wave $\varphi
_\text{in}(x_\text{in}^{+})$ in  Region I, the experiments determine
whether nonlinear interactions of the edge waves are generated in  Region II. If the amplitude of the output wave $\varphi_\text{ out}(x_\text{out}^{+})$
does not increase linearly with the increase in the amplitude of $\varphi _\text{in}(x_\text{in}^{+})$, the nonlinear interactions exist. If the experiments show a
breakdown of the edge unitarity conditions of $\alpha (k,k^{\prime })$ and $\beta
(k,k^{\prime })$ in Eq.~(\ref{e24}) and Eq.~(\ref{e25}), the edge excitations
interact with bulk excitations. 
More detailed analyses will be reported in forthcoming papers.

We have also shown that the expanding edges can be regarded as analogs of expanding universes in 2D
dilaton gravity models, including the JT model and dSJT model. By controlling the
time schedule of the edge expansion, the experiments are capable of
simulating the gravity models with the potential term $V\left( \Phi \right) $ in Eq.~(\ref{30}). Furthermore, we 
demonstrated the analog Hawking radiation from the de Sitter horizon formed in the expanding edge region, which might be detected in our experimental setup. Since the field $\varphi$ in Eq.~(\ref{e22}) for a QH system in a low-noise situation is regarded as a quantum field in an expanding analog universe, the time-dependent QH systems enable
us to explore the trans-Planckian problem and the quantum-classical
transition problem in 2D dilaton gravity.

In conclusion, future studies of the expanding edge experiments are expected to reveal new
physics of QH systems, yielding numerous milestones that endeavour to resolve the fundamental problems in early universe.

\begin{acknowledgments}
The authors are grateful to K. Nakayama, T. Nishioka, N. Shibata, T. Shiromizu, T. Takayanagi, M. Tezuka, and K. Yonekura for their fruitful discussions. This work was supported by a Grant-in-Aid for Scientific Research (Grant Nos. 21H05188, 21H05182 (M. H., G. Y.), JP19K03838 (M. H.), 17H01037 (G. Y.), 19H05603 (G. Y.), 21F21016 (G. Y.), and 19K03866 (Y. N.)) from the Ministry of Education, Culture, Sports, Science, and Technology (MEXT), Japan.
\end{acknowledgments}

\appendix*

\section{Formula for 2D gravity}

\begin{itemize}
\item Flat spacetime metric:%
\begin{equation*}
ds^{2}=-c^{2}dt^{2}+dx^{2}.
\end{equation*}

\item Definition of Christoffel symbols for general metric form $%
ds^{2}=g_{\mu \nu }dx^{\mu }dx^{\nu }$: 
\begin{equation*}
\Gamma _{\beta \gamma }^{\alpha }=\frac{1}{2}\,g^{\alpha \mu }\left( \partial
_{\beta }g_{\mu \gamma }+\partial _{\gamma }g_{\mu \beta }-\partial _{\mu
}g_{\beta \gamma }\right) .
\end{equation*}

\item Definition of Riemann curvature tensor:%
\begin{eqnarray*}
R^{\alpha }{}_{\beta \mu \nu } &=&\partial _{\mu }\Gamma _{\nu \beta }^{\alpha
}-\partial _{\nu }\Gamma _{\mu \beta }^{\alpha }+\Gamma _{\mu \gamma
}^{\alpha }\Gamma _{\nu \beta }^{\gamma }-\Gamma _{\nu \gamma }^{\alpha
}\Gamma _{\mu \beta }^{\gamma }, \\
R_{\alpha \beta \mu \nu } &=&g_{\alpha \gamma} R^{\gamma}{}_{\beta \mu \nu
}
\end{eqnarray*}

\item Definition of Ricci curvature tensor:%
\begin{equation*}
R_{\alpha \beta }=R^{\mu }{}_{\alpha \mu \beta }=g^{\mu \nu }R_{\mu \alpha \nu
\beta }.
\end{equation*}

\item Definition of scalar curvature:%
\begin{equation*}
R=g^{\alpha \beta }R^{\mu }{}_{\alpha \mu \beta }=g^{\alpha \beta }g^{\mu \nu
}R_{\mu \alpha \nu \beta }.
\end{equation*}

\item In 2D gravity theory, $R_{\alpha \beta \mu \nu }$ and $R_{\alpha \beta
}$ are uniquely determined by $R$ and $g_{\mu \nu }$ as 
\begin{eqnarray*}
R_{\alpha \beta \mu \nu } &=&\frac{1}{2}R\left( g_{\alpha \mu }g_{\beta \nu
}-g_{\alpha \nu }g_{\beta \mu }\right) , \\
R_{\alpha \beta } &=&\frac{1}{2}\,g_{\alpha \beta }R.
\end{eqnarray*}

\item Light cone coordinate systems:%
\begin{equation*}
x^{\pm }=c\,t\pm x=x^{0}\pm x^{1}.
\end{equation*}

\item Partial derivatives with respect to $x^{\pm }$: 
\begin{eqnarray*}
\partial _{+} &:&=\frac{\partial }{\partial x^{+}}=\frac{\partial x^{0}}{%
\partial x^{+}}\,\partial _{0}+\frac{\partial x^{1}}{\partial x^{+}}\,\partial
_{1}=\frac{1}{2}(\partial _{0}+\partial _{1}), \\
\partial _{-} &:&=\frac{\partial }{\partial x^{-}}=\frac{\partial x^{0}}{%
\partial x^{-}}\,\partial _{0}+\frac{\partial x^{1}}{\partial x^{-}}\,\partial
_{1}=\frac{1}{2}(\partial _{0}-\partial _{1}).
\end{eqnarray*}

\item Any metric form can be rearranged into a conformally flat metric form as%
\begin{equation*}
ds^{2}=-\exp \left( 2\Theta (x^{+},x^{-})\right) dx^{+}dx^{-},
\end{equation*}%
at least, in any local region of whole spacetime.

\item Nonvanishing $\Gamma _{\beta \gamma }^{\alpha }$ in conformally flat
coordinate system: 
\begin{eqnarray*}
\Gamma _{++}^{+} &=&2\,\partial _{+}\Theta , \\
\Gamma _{--}^{-} &=&2\,\partial _{-}\Theta .
\end{eqnarray*}

\item Scalar curvature in conformally flat coordinate system: 
\begin{equation*}
R=8\,e^{-2\Theta }\,\partial _{+}\partial _{-}\Theta .
\end{equation*}
\end{itemize}

\end{document}